\documentclass[%
pre,
amsmath,amssymb,
reprint,%
]{revtex4-1}

\usepackage{graphicx}
\usepackage{dcolumn}
\usepackage{bm}

\usepackage[utf8]{inputenc}
\usepackage[T1]{fontenc}
\usepackage{mathptmx}

\begin{document}
	
	\preprint{AIP/123-QED}
	
	\title{Design of conditions for self-replication}
	
	\author{Sumantra Sarkar}
	\altaffiliation[]{Center for Nonlinear Studies, Los Alamos National Laboratory, Los Alamos, NM 87544, USA}
	\email{sumantra@lanl.gov}
	\author{Jeremy L. England}%
	\email{jengland@mit.edu}
	\affiliation{ 
		Physics of Living Systems, Massachusetts Institute of Technology, 400 Technology Square, Cambridge, MA 02139, USA
	}%

	\date{\today}
	
	\begin{abstract}
		A ``self-replicator" is usually understood to be an object of definite form that promotes the conversion of materials in its environment into a nearly identical copy of itself.  The challenge of engineering novel, micro- or nano-scale self-replicators has attracted keen interest in recent years, both because exponential amplification is an attractive method for generating high yields of specific products, and also because self-reproducing entities have the potential to be optimized or adapted through rounds of iterative selection.  Substantial steps forward have been achieved both in the engineering of particular self-replicating molecules, and also in characterizing the physical basis for possible mechanisms of self-replication.  At present, however, there is need for a theoretical treatment of what physical conditions are most conducive to the emergence of novel self-replicating structures from a reservoir of building blocks on a desired time-scale.  Here we report progress in addressing this need.  By analyzing the kinetics of a toy chemical model, we demonstrate that the emergence of self-replication can be controlled by coarse, tunable features of the chemical system, such as the fraction of fast reactions or the width of the rate constant distribution. We also find that the typical mechanism is dominated by the cooperation of multiple interconnected reaction cycles as opposed to a single isolated cycle. The quantitative treatment presented here may prove useful for designing novel self-replicating chemical systems.
	\end{abstract}
	
	\maketitle

	
	\section{Introduction}
	Emergence of self-replicators from a mixture of components is marked by exponential growth of one or more multi-component structures. This process is of great practical importance due to the possibility of exponentially fast synthesis of target structures, and also has previously been considered in models of pre-biotic chemistry~\cite{butlerow1861formation,breslow1959mechanism,dyson1982model,szathmary2006origin,king1982recycling}. The mechanisms that enable self-replication in a soup of metastable bound states have been investigated intensively in the past decades~\cite{bissette2013mechanisms,paul2004minimal} and still continue to inspire new attempts \cite{carnall2010mechanosensitive, wang2011self,vaidya2012spontaneous,zeravcic2014self, sadownik2016diversification,vsaric2016physical,barenholz2017design,zwicker2017growth, england2013statistical,perunov2016statistical}. The processes of self-replication described in these studies, though distinct, share two mechanistic elements: (a) the existence of at least one autocatalytic cycle and (b) a source of driving that runs the autocatalytic cycle.
	
\begin{figure*}[htbp]
		\centering
		\includegraphics[width=17cm]{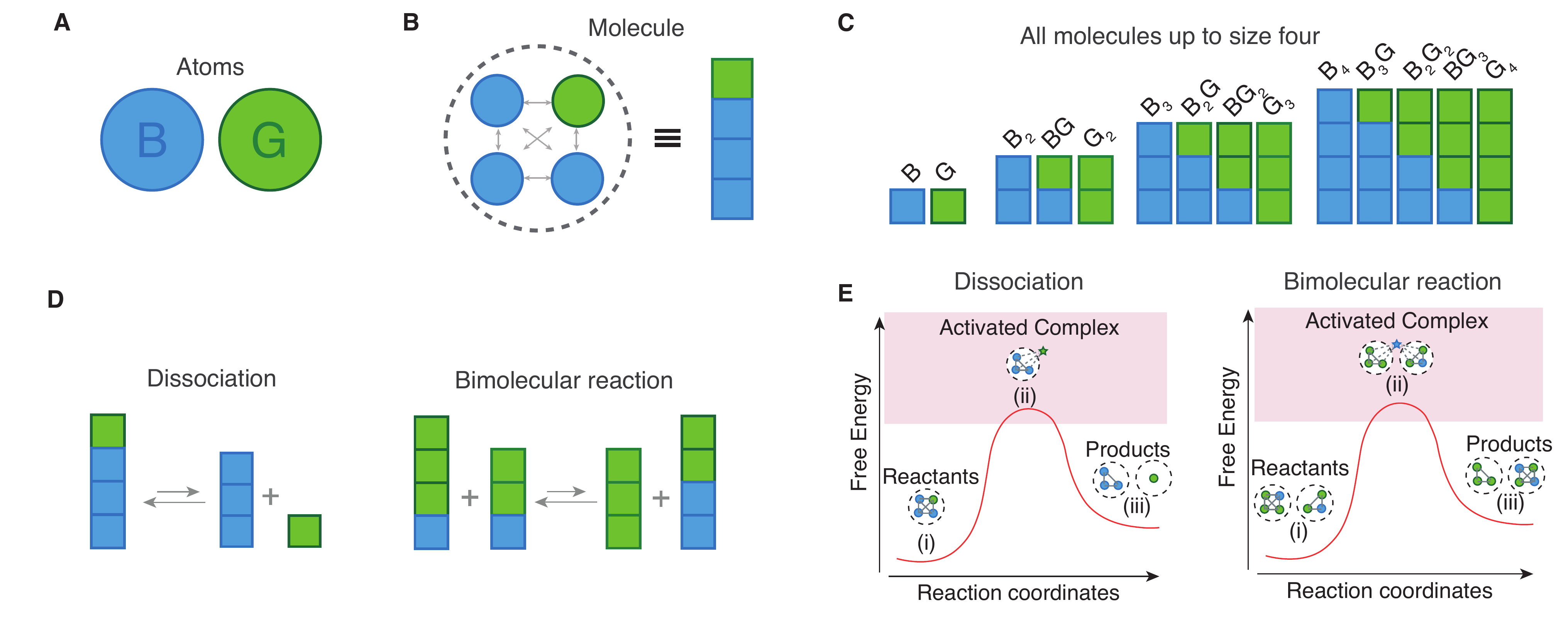}
		\caption{\textbf{Toy chemical system with two monomer types: } (\textbf{A}) The toy chemical system (\textit{see Materials \& Methods for details}) consists of two \textit{atoms}, $ B $ and $ G $. (\textbf{B}) The atoms interact with each other with three interaction energies $ \epsilon_{BB}, \epsilon_{BG}, \epsilon_{GG}$ (measured in units of $ kT $) to form \textit{molecules}, which are represented as stack of atoms. $ \epsilon_{XY} $ denotes the interaction energy between an atom of type $ X $ and another atom of type $ Y $. In the figure, the gray arrows represent interaction between different atoms. All atoms inside a molecule, such as the one shown here, interact with each other. (\textbf{C}) The current model consists of fourteen molecules that  contains at most four atoms. (\textbf{D}) The molecules take part in dissociation or bimolecular reactions. (\textbf{E}) In the mechanistic model (see Materials \& Methods), the rate constants of these reactions are calculated from a transition state model. The star shaped atoms are the atoms that are being donated. Circular atoms are other atoms in the molecules.}
		\label{fig1}
	\end{figure*}

	In the usual case~\cite{bissette2013mechanisms} an autocatalytic cycle is designed by experimenters to consume one or more building blocks that are provided in excess to generate replicas of a template that is used as a seed. A significant challenge in any such case lies in devising an appropriate chemical library that limits parasitic side reactions. Theoretical approaches, meanwhile, have been most successful in the opposite regime, where the catalytic network is sufficiently densely connected, and every molecule available in the reaction pot catalyzes the production of at least one other molecule~\cite{kauffman1986autocatalytic,jain2001model, zeravcic2017spontaneous}. In such a case, it is possible to formulate general criteria for the onset of positive feedback loops in the catalytic reaction network that lead to the exponential growth of the molecules in those loops. Thus, although it is qualitatively understood that robust self-replication requires sufficient catalytic promiscuity that somehow avoids excessive side reactions, there is need for a quantitative treatment of this tradeoff in a physical model that may provide future guidance for the design of conditions conducive to the spontaneous emergence of self-replicators from customizable mixtures of nano- or microscale components \cite{wang2012colloids,zhang2014self}.  Therefore, we sought to investigate a toy model where all possible stoichiometric combinations of certain building blocks are considered in the construction of an effective model of a ``chemical" space . Using this model (Fig.~\ref{fig1}), we lay out general conditions for the emergence of exponential growth in systems without explicit catalysis.  Interestingly, we find that the typical mechanism for the emergence of self-replicators occurs via a multi-cycle topological element in the reaction, and therefore violates previously established quantitative criteria for self-replication that were developed assuming that self-replication occurs through isolated autocatalytic cycles \cite{szathmary2006origin,king1982recycling}.   
	
	\section{Model}
	
	\subsection{Toy chemical system}
	We undertook to model a large, well-mixed reaction pot with diverse possible combinations of monomers.  We call these monomers ``atoms" here because we eventually plan to model the dynamics of their bound states using thermodynamically consistent mass-action kinetics, but it should not be imagined that we intend exclusively or even principally to describe real molecular chemistry using the model presented here.  Rather, the essence of the ``chemical space" constructed is that it is a vast space of diverse combinations among physical interacting components such as polymer-coated colloidal particles or DNA origami  (Fig.~\ref{fig1}A).
	
	In our model, two or more atoms interact with each other to form a bound state, which we call a ``molecule." For simplicity, we assume that the molecules do not have any internal structure and all the atoms inside a molecule interact with all other atoms in that molecule with interaction energies $\epsilon_{BB}, \epsilon_{BG}, \mbox{ or } \epsilon_{GG}$ (Fig.~\ref{fig1} B).  Since the molecules do not have any internal structure, their free energies are completely determined by their composition and the three $\epsilon$ parameters. Also, we assume that each molecule contains at most $ \mu_{max} $ atoms, and forbid all other bound states. Except where it is explicitly mentioned, we set $ \mu_{max} = 4 $.  With these two assumptions it can be shown that there are fourteen distinct molecules in the model with two types of monomers(Fig.~\ref{fig1} C) .
	
	The molecules take part in reactions that involve one molecule donating an atom to the surrounding medium or to another molecule. We call the former a dissociation reaction and the latter a bimolecular reaction (Fig.~\ref{fig1} D). The reactions are activated processes and the rate constant of a given reaction that takes the reactant state $ i $  to product state $ j $, is inversely proportional to the exponential of the barrier height: $ k_{ij} \propto  \exp(- B_{ij}) $. The activation barriers $ B_{ij} $ are either chosen randomly or using a model of the transition state. We refer to the latter as \textit{mechanistic model}.
	
	In the mechanistic model, $ B_{ij}  = F^{Tr}_{ij} - F_i$, where $ F_i $ is the free energy of the reactant state and $ F^{Tr}_{ij} $ is the free energy of the transition state. $ F_i $ is determined from the interaction energies. To calculate $ F^{Tr}_{ij} $, we assume that during a reaction, the donated atom first goes to an excited state, where it interacts with other atoms in the donor molecule through a weakly repulsive interaction (Fig.~\ref{fig1} E) that is proportional to the ground state interaction energy. The proportionality factor $ c_0 = -0.1$ is same for all three interaction energies and is a parameter of the model. The results described is robust with variation in $ c_0 $, as long as $\epsilon_{**} < 0 \mbox{ and } c_0 < 0 $.

	The resulting toy ``chemistry" generates a full system of rate equations with mass-action kinetics governing the concentrations of different allowed molecules. There is no explicit catalysis or autocatalysis in this system at the level of a single reaction, but catalytic and autocatalytic cycles appear naturally in the reaction network (defined in the next section) due to coupling between different reactions. In what follows, we explicitly solve this set of equations in two instances of the model with one and two types of atoms. We investigate the resultant transient kinetics of molecular concentration to identify conditions necessary for the persistence of one or more autocatalytic cycles that drive exponential growth of a subset of the molecules.

	\subsection{Reaction network}
	\begin{figure*}[htbp]
		\centering
		\includegraphics[width=17cm]{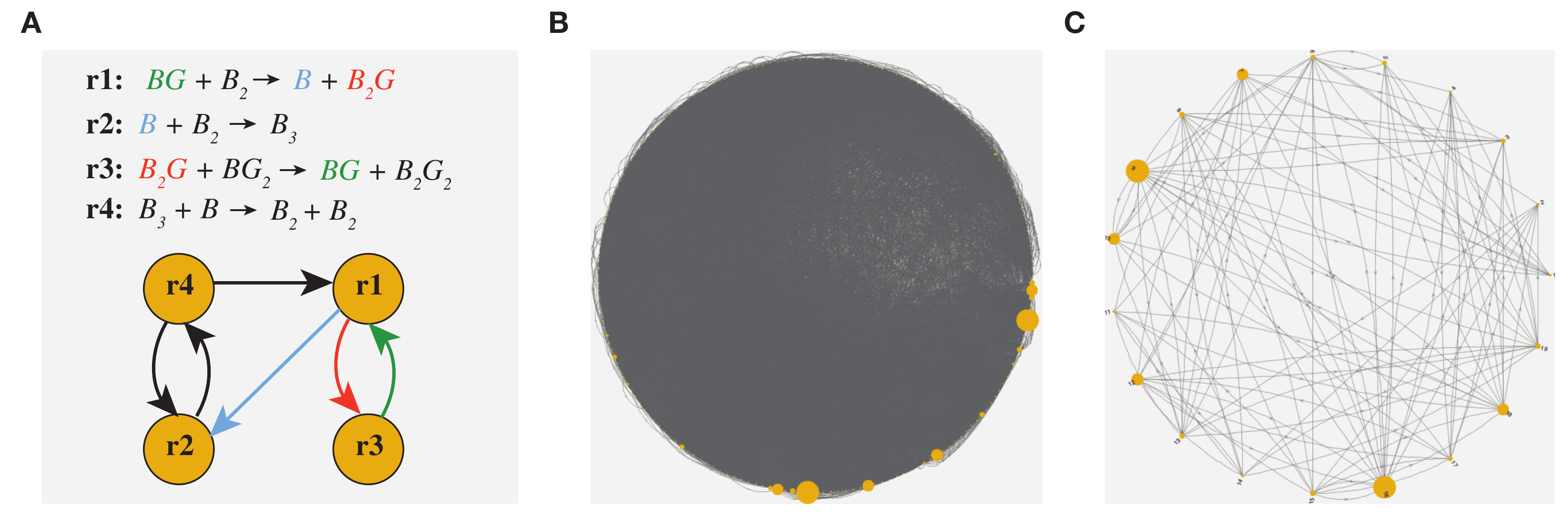}
		\caption{\textbf{Reaction Network:}(A) \textit{The definition of the reaction network:} The reactions are the nodes of this network and a directed edge from node $ \textbf{ri}$ to $ \textbf{rj} $ exists, if any of the product of reaction $ i $ is used as the reactant in reaction $ j $. For example, in the example considered here, \textbf{r1} produces $ B $ and $ B_2G $, which are used as reactants in \textbf{r2} and \textbf{r3} respectively. Hence, as shown, in the reaction network there are directed edges from \textbf{r1} to \textbf{r2} and \textbf{r3}. Similarly, $ BG $ is a product of \textbf{r3}, which is used as reactant in \textbf{r1}. Hence, there is a directed edge from \textbf{r3} to \textbf{r1}. (B) The generic structure of the reaction network of our model. The size of the nodes are proportional to the reaction rate. As can be seen, the reaction network is very dense, but only very few reactions contribute significantly to the instantaneous kinetics. Consequently, (C) the reaction kinetics is effectively determined by a sparser reaction network consisting of the reactions with rates greater than a threshold value. We choose this threshold value to be $ 10\% $ of the maximum rate. }
		\label{fig:rnet}
	\end{figure*}
	\subsubsection{Coupled reaction graph}
	
	In our model, the product of various reactions acts as reactants to other reactions. For example in the following two reactions one of the products of \textbf{r1}, $ B $ is used as a reactant in \textbf{r2}.
	\begin{eqnarray}
	\textbf{r1:}& &BG + B_2 \rightarrow B + B_2G \label{r1}\\
	\textbf{r2:}& &B + B_2 \rightarrow B_3 \label{r2}
	\end{eqnarray}
	Hence, \textbf{r1} is coupled to \textbf{r2}. We graphically represent this relationship by constructing a directed graph, whose nodes are the reactions \textbf{r1} and \textbf{r2} and which has a directed edge from \textbf{r1} to \textbf{r2} (Fig.~\ref{fig:rnet}A). The graphical representation of all the 180 reactions in our model is shown in Fig.~\ref{fig:rnet}B. Three reaction motifs are usually found in the reaction network: catalytic cycles, autocatalytic cycles, and lossy side reactions.
	
	\subsubsection{Network motifs}
	\paragraph{Catalytic cycles (CC)}
	Consider the reactions \textbf{r1} and \textbf{r3} in Fig.~\ref{fig:rnet}A. Both of them have a directed edge from one to the other. Hence, if by some process \textbf{r1} and \textbf{r3} runs in sequence for some time, then the net output will be the production of $ B $ and $B_2G_2 $ from $ B_2 $ and $ BG_2 $, catalyzed by $ BG $ and $ B_2G $. It is easy to show that other cycles, such as $ \textbf{r1}\rightarrow \textbf{r2}\rightarrow \textbf{r4} \rightarrow \textbf{r1} $ and  $ \textbf{r2}\rightarrow\textbf{r4} \rightarrow \textbf{r2} $ are also catalytic cycles. In fact, any cycle in the reaction graph defined here is a catalytic cycle.
	
	\paragraph{Autocatalytic cycles (ACC)}
	A subset of the catalytic cycles have a special property that at least one of the catalyst molecules is produced in excess. That is the catalyst molecule catalyzes its own production. We refer to such cycles as autocatalytic cycles. For example, it is easy to see that $ \textbf{r2}\rightarrow\textbf{r4} \rightarrow \textbf{r2} $ is an autocatalytic cycle, because $ B_2 $ catalyzes its own production.
	
	\paragraph{Lossy side reactions}
	In a complex reaction network, such as ours, it is likely that reactions are coupled to more than one reactions. Therefore, quite often, the function of an autocatalytic cycle is hindered by the presence of parasitic side reactions that couple to one of the reactions in the autocatalytic cycle and usurp the resources required to drive the cycle. For example, \textbf{r1} is a lossy side reaction for the autocatalytic cycle $ \textbf{r2}\rightarrow\textbf{r4} \rightarrow \textbf{r2} $. As can be seen in Fig.~\ref{fig:rnet}A, lossy reactions need not be an isolated reaction. Often, they are part of another catalytic or autocatalytic cycles. When it is part of another autocatalytic cycle, the parasitism is equivalent to competition between two autocatalytic cycle.
	
	\section{Conditions for self-replication}
	The physico-chemical conditions required for self-replication is very different in an interacting chemical system, such as ours, than for isolated autocatalytic cycles which have been studied theoretically and experimentally over the last few decades. Prior work has indicated that the kinetic dominance of reactions can be quantified through a measure called specificity. It has been shown that for any cycle, the product of the specificity, which we call cycle-specificity for the sake of brevity, has to be greater than 0.5 for a reaction cycle to run. However, this result is incomplete. As we show here, even for an isolated autocatalytic cycle, other conditions have to be met for self-replication to take place. Furthermore, self-replication in an interacting system can happen even when the cycle-specificity of all the autocatalytic cycles is orders of magnitude lesser than 0.5, requiring a fresh search for the conditions required for self-replication.
	
	\begin{figure}[htbp]
		\centering
		\includegraphics{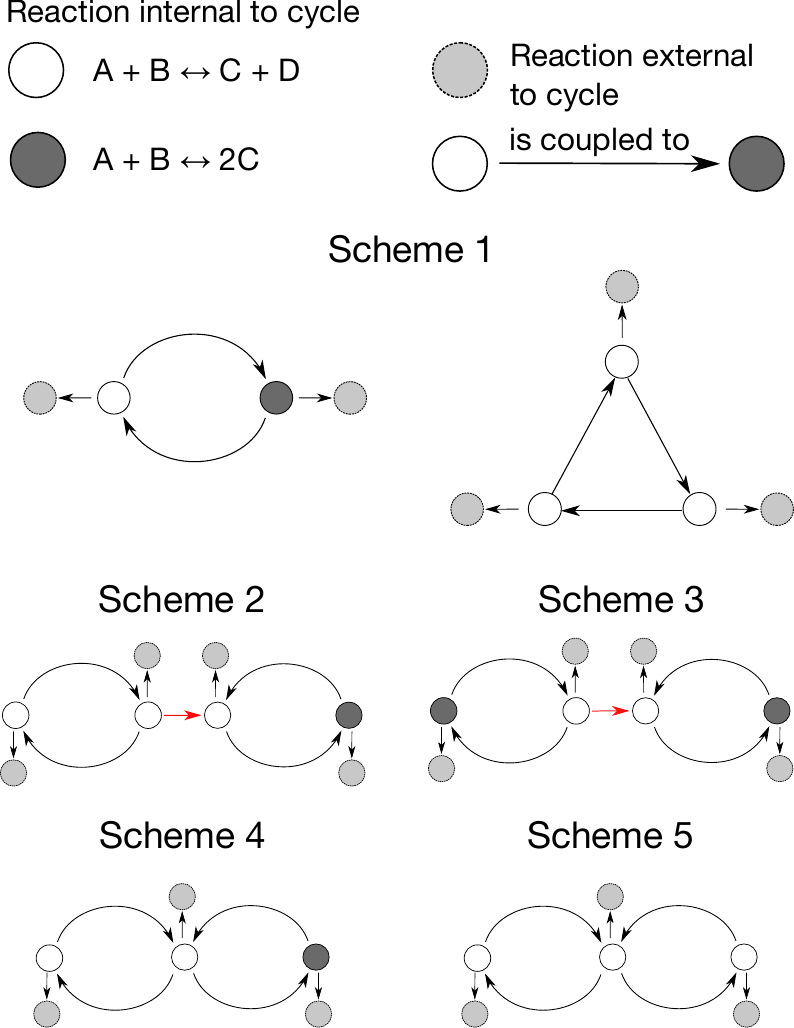}
		\caption{\textbf{Modes of self-replication:} \textit{Scheme 1:}  Isolated autocatalytic cycles (ACC). On the left is a two step ACC and on the right is a three step ACC. \textit{Scheme 2:} A catalytic cycle (CC) is coupled to an ACC through the waste product of the former (red arrow). \textit{Scheme 3:} ACC coupled to another ACC through the waste of the former. \textit{Scheme 4:} CC is coupled to ACC through a catalyst, which is also catalyst for the ACC. \textit{Scheme 5:} CC is coupled to another CC by sharing a catalyst molecule between them. In all of these schemes the light gray reactions are reactions that couple to the reactions in a motif, but are not part of it.}
		\label{fig:motifs}
	\end{figure}
	
	To establish these conditions, we study kinetics of simple network motifs that are outlined in Fig.~\ref{fig:motifs}. These are by no means the exhaustive list of network motifs that lead to self-replication, but these are the simplest ones to study. We summarize the necessary conditions for self-replication for these motifs below. The derivation of these condition is described in SI. The sufficient condition for self-replication is the union of all the necessary conditions.
	
	\paragraph{Scheme 1: Isolated ACC} For isolated ACCs, the cycle-specificity has to be greater than 0.5, in agreement with previous results. However, additionally, the chemical current (see Materials \& Methods for definition) for all the reactions have to be greater than zero and increasing function of time.
	
	\paragraph{Scheme 2 and 3:} For scheme 2, no exponential growth occurs unless the specificity of ACC is greater than 0.5. For scheme 3, it is possible to observe exponential growth as long as one of the ACC has specificity greater than 0.5.
	
	\paragraph{Scheme 4 and 5:} It is difficult to write a simple closed expression for the condition required for exponential growth. However, under these two schemes, it is possible to observe exponential growth even when both cycles have specificity lower than 0.5. The specificity distribution required for these two schemes is listed in Table~\ref{tab:spec}.
	\begin{table}
		\centering
		\begin{tabular}{|c|c|c|c|c|}
			\hline
			$\sigma_1$ & \multicolumn{2}{|c|}{$\sigma_2$ } &\multicolumn{2}{|c|}{$\sigma_3$ } \\
			\hline
			& Scheme 4 & Scheme 5 & Scheme 4 & Scheme 5 \\
			\hline
			0.1 & 0.8418 & 0.9180 & 0.8154 & 0.9180 \\
			0.2 & 0.7356 & 0.8542 & 0.6142 & 0.8294 \\
			0.3 & 0.6573 & 0.7780 & 0.4422 & 0.7443 \\
			0.4 & 0.5693 & 0.7356 & 0.2130 & 0.6487 \\
			0.5 & 0.5232 & 0.6817 & 0.0318 & 0.5509 \\
			0.6 & 0.4768 & 0.6573 & 0.0001 & 0.4409 \\
			0.7 & 0.4307 & 0.6142 & 0.0001 & 0.3427 \\
			0.8 & 0.4007 & 0.5693 & 0.0001 & 0.2413 \\
			0.9 & 0.3858 & 0.5693 & 0.0001 & 0.1241 \\
			1.0 & 0.3427 & 0.5232 & 0.0001 & 0.0002\\
			\hline
		\end{tabular}
		\vspace{0.5cm}
		\caption{Conditions for self-replication for Schemes 4 and 5. The value tabulated are the numerically obtained minimum specificity required for the concentration to grow exponentially. 0.0001 is the minimum value numerically investigated. Therefore, it is possible that for entries with tabulated value 0.0001, the minimum specificity required may be lesser than 0.0001. $\sigma_1$ is the specificity of the reaction that is shared by both cycles. $\sigma_{2,3}$ are the other reactions in the scheme. Particularly for Scheme 4, $\sigma_3$ denote the specificity of the doubling reaction (dark circle in Fig.~\ref{fig:motifs}).}
		\label{tab:spec}
		
	\end{table}

	\section{Coarse control of exponential growth}
	
	The fundamental goal of this paper is to understand how these reaction motifs come to dominate the kinetics and give rise to different types of concentration growth. For example, if the kinetics is dominated by autocatalytic cycles, we expect to observe exponential growth, whereas if the the kinetics is dominated by uncoupled reactions, then we expect linear growth. It is to be noted that growth is a strictly transient behavior of the underlying rate equations, which is governed by the topology of the coupled reaction graph and the instantaneous rates of the reactions. Therefore, through a suitable choice of reaction library, which determines the topology, and rate constants, which determine the instantaneous rates, it is possible to manipulate the influence of various motifs on the reaction kinetics.
	
	These facts are well known and have been used qualitatively to design small chemical systems that permit near-exponential growth of molecular concentrations~\cite{bissette2013mechanisms}. However, such qualitative knowledge is of little use when large chemical systems with hundreds, if not thousands, of reactions need to be designed for self-replication. To design a chemical network of such complexity, quantitative relationship between the rate constants and the transient behavior of the reaction network need to be established. Unfortunately, it is impractical to explore the parameter space of the rate constants to establish such a relationship due to the cost involved with exploring the parameter space, which may be thousand dimensional. We therefore need to establish the required quantitative behavior using coarse (macroscopic) features of the rate constants, for example, in increasing order of coarseness, (a) \textbf{Protocol PF: } the fraction of fast reactions, (b) \textbf{Protocol CD: } the width of the rate constant distribution, or (c) \textbf{Protocol IE: } the interaction energies between the atoms. Due to our interest in self-replication, we only focus on the emergence of exponential growth and establish quantitative criteria using these parameters.
	
	\begin{figure}[htbp]
		\centering
		\includegraphics[width=8.5cm]{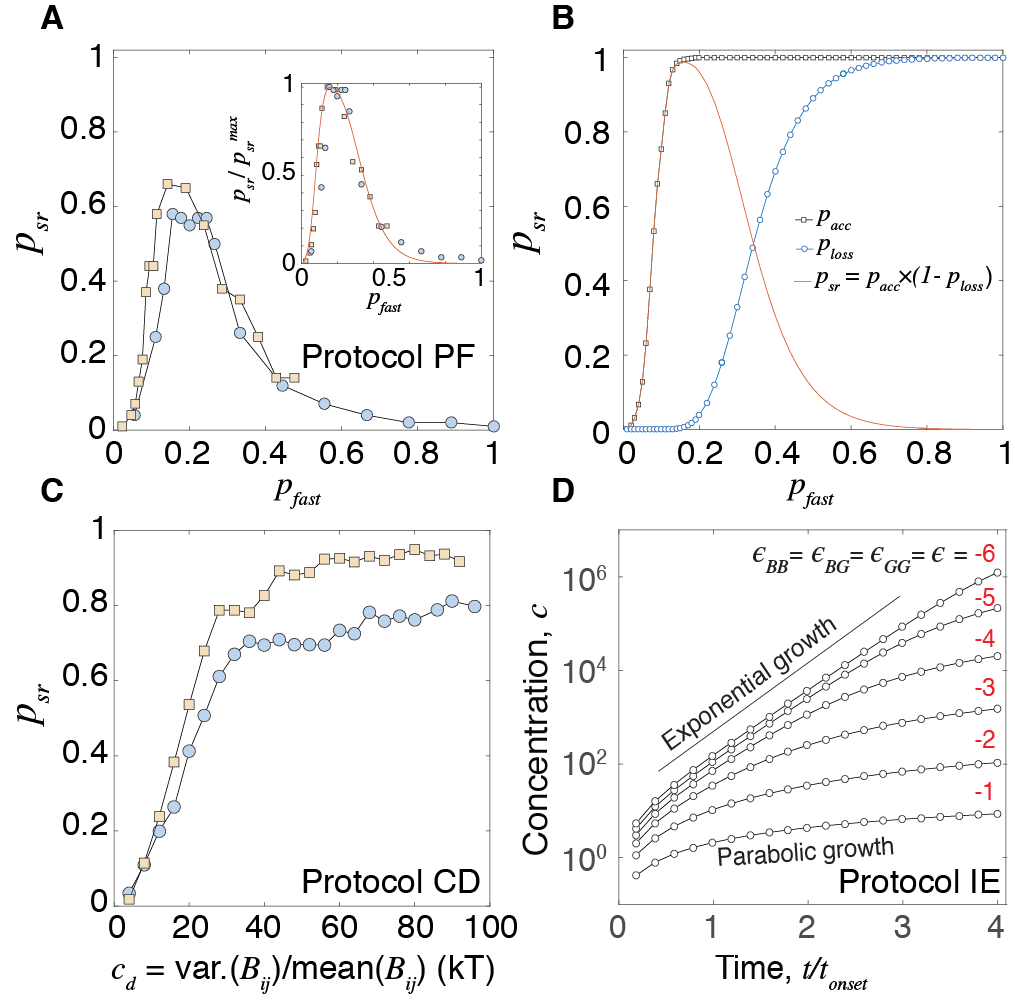}
		\caption{\textbf{Coarse control of exponential growth:} (A) \textbf{Protocol PF:} Numerical solution of the rate equations show that the probability of self-replication, $p_{sr}$ is maximum for an intermediate value of fraction of fast reactions, $ p_{fast} $. Results are shown for $ \mu_{{max}}  = 4 $ (blue circle) and $ 5 $ (orange square), where $ \mu_{\mbox{max}} $ is the maximum allowed size of the molecules. \textit{Inset:} Analytical prediction of $ p_{sr} $ (shown in B) normalized by its maximum value vs. $ p_{fast} $  (orange curve) matches well with numerical data. (B) Probability of finding a fueled autocatalytic cycle, $ p_{acc}$ (gray square) and the probability of loss of fuel due to side reaction, $ p_{loss}$ (blue circle) as a function of the probability of fast reactions, $ p_{fast}$. Probability of self replication, $ p_ {sr}  = p_ {acc}\times \left(1 - p_{loss}\right)$, is plotted in orange line.(C) \textit{Protocol CD:} $ p_{sr} $ vs the coefficient of dispersion (variance/mean) of the distribution of activation barriers, $ B_{ij} $ for  $ \mu_{max} = 4, t_{obs} = \infty $ (blue circle) and $ \mu_{max} = 5, t_{obs} = \infty $ (orange square). For narrow distribution ($< 10 kT$), no exponential growth is observed and only power law growth is observed (SI). For broader ($> 10 kT$), $ p_{sr} $ increases and eventually saturates with the dispersion of the activation barriers.  (D) \textbf{Protocol IE:} As the magnitude of interaction energy increases, the concentration tends to grow exponentially. For example, for $ \epsilon_{BB,BG,GG} = -1 $, parabolic growth is observed. However, for $ \epsilon{BB,BG,GG} = -6 $, exponential growth is observed.}
		\label{fig:protocols}
	\end{figure}
	
	\subsection{PF: Fraction of fast reactions}
	The most theoretically accessible case arises when all the interaction energies are zero and the rate constants are chosen in such a way that a controllable fraction, $ p_{fast} $, of the reactions may occur, and the rest are effectively forbidden. To implement such a system, we identified the set of all reactions permitted by stoichiometry and drew the random barriers for the reactions from the binary set $ \{0, \infty\} $, corresponding to rate constants of 1 or 0. The fast reactions, with rate constants $ 1 $, were assigned with a probability $ p_{fast} $. To ensure detailed balance conditions, the barriers for the forward and the reverse reactions were set to be equal. As we discuss later in this section, $ p_{fast} $ can be mapped to the dispersion of the rate constant distribution, with $ p_{fast} \approx 1 $ corresponding to narrow and $ p_{fast} \approx 0$ to broad distributions. 
	
	Under these assumptions, the probability of self-replication, $ p_{sr} $, can be estimated (SI) as a function of $ p_{fast} $. Self-replication occurs if and only if at least one autocatalytic cycle in the reaction network has direct and exclusive access to its fuel (Fig.~\ref{fig:protocols}A). Hence, $ p_{sr} $ can be calculated from (a) the probability of finding at least one autocatalytic cycle with direct access to its fuel, $ p_{acc}(p_{fast}) $ and (b) the probability that all autocatalytic cycles have side reactions, $ p_{loss}(p_{fast}) $. Whence, for $ p_{fast} = x $:
	\begin{equation}
	p_{sr}(x) = p_{acc}(x)\times\left(1 - p_{loss}(x)\right). \label{eqn:psr}
	\end{equation}
	As Fig.~\ref{fig:protocols}A-B shows, self-replication generally sets in spontaneously when a reaction network has a specific level of complexity dictated by the trade off of the two different competing percolation transitions, $ p_{acc} $ and $ p_{loss} $ -- the first of which determines whether there are enough fast reactions to ensure existence of at least one driven autocatalytic cycle, and the second of which determines whether reactions are so promiscuously coupled that every cycle is drained by numerous side reactions. Due to this trade off, an optimal $ p_{fast} $ exists at which $ p_{sr} $ is maximized. Simply stated, this result implies that emergent self-replication occurs with high probability when there are enough autocatalytic cycles and no parasitic reactions: a result that is qualitatively well-known~\cite{bissette2013mechanisms} and perhaps unsurprising. More surprisingly, however, our quantitative treatment shows that this optimality depends only on the reaction network topology (through $ p_{fast} $ and the randomized graph connectivity) and should be relatively insensitive to the specific rate constant distribution. Therefore, as long as $ p_{fast} $ can be tuned to its optimal value, exponential growth will emerge in a large network with certainty. \textit{What remains now is to determine whether a quasi-randomly connected network is a suitable approximation to real chemical network, and if so, how then may we tune the effective value of $ p_{fast} $ to its optimal value}?
	
	\subsection{CD: Width of the rate constant distribution}
	
	A first and simplest hypothesis is that the $ p_{fast} $ can be tuned to optimality by the dispersion of the rate constants. To demonstrate this, we chose the activation barriers from exponential distributions with varying amount of coefficient of dispersion (variance/mean), $ c_d $, while keeping the interaction energy zero. In the first set of studies, we numerically solved the equations until concentrations reached steady state ($ t_{obs} =\infty $). From the obtained time-series of the molecular concentrations, we found their growth exponent $ \gamma $ (M\&M). If $ \gamma = 1 $, the corresponding concentration grows exponentially. If $ \gamma < 1 $, the concentration grows subexponentially. The probability of exponential growth, $p_{sr} $, was determined by finding $Prob(\gamma > 0.99)$. Under this protocol, when the distribution was too narrow ( $ c_d < 10 kT $ in Fig.~\ref{fig:protocols}C), the molecules never grew exponentially. However, when the distribution was broader, the probability of exponential growth, $ p_{sr}$, increased with $ c_d $, eventually saturating at a value that is dependent on the underlying reaction network (Fig.~\ref{fig:protocols}C). 
	

	
	\subsection{IE: Interaction energy}
	
\begin{figure}
		\centering
		\includegraphics[width=8.5cm]{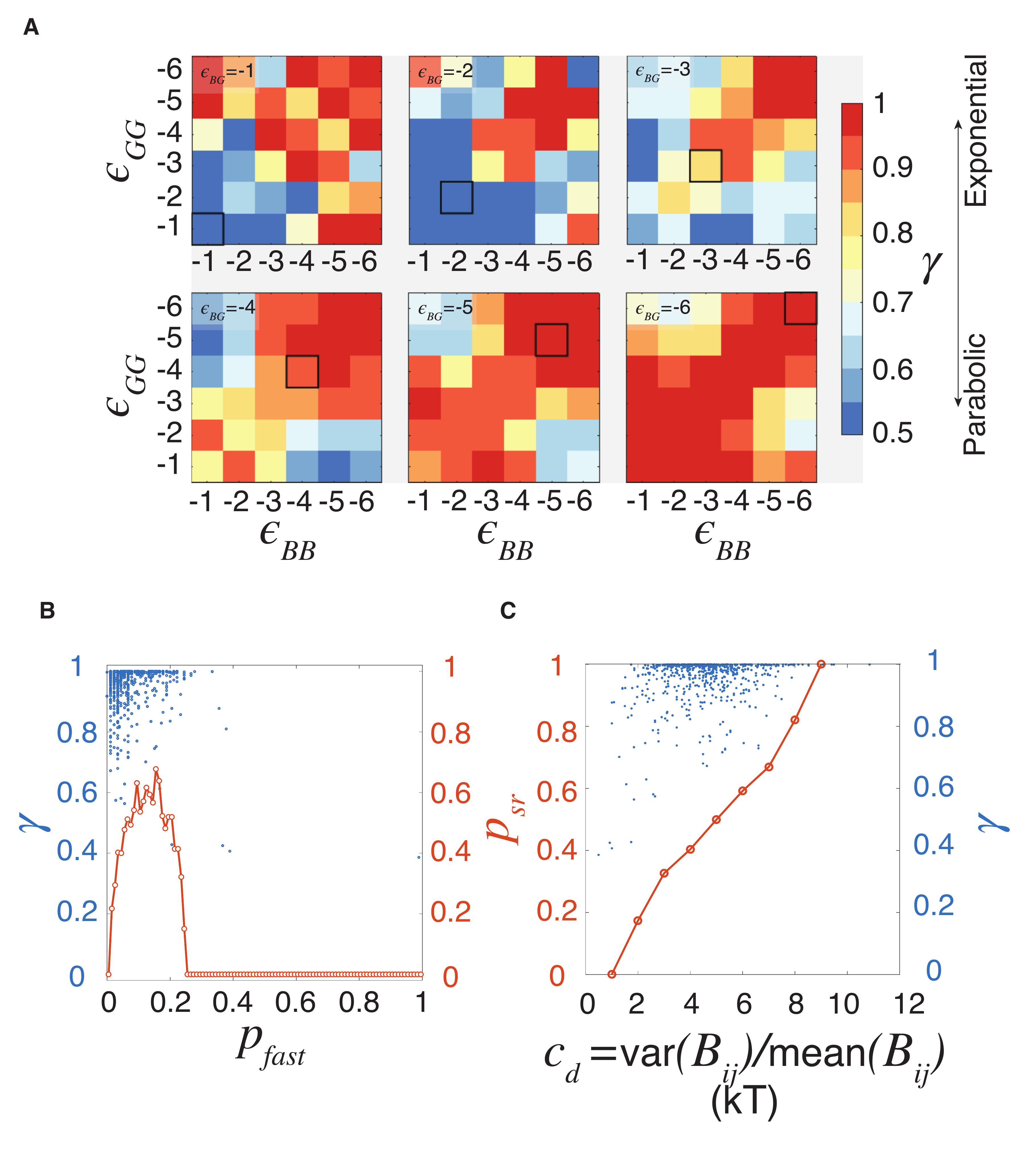}
		\caption{\textbf{Protocol IE:}(A) Growth exponent $ \gamma $ for different combination of interaction energies. Red correspond to $ \gamma = 1 $ (exponential growth) and blue correspond to $ \gamma = 0.5 $ (parabolic growth). (B) $ \gamma $ as a function of an estimate of fraction of fast reaction (see M\&M), $ p_{fast} $. Probability of self-replication $ p_{sr} $ is defined as probability of finding $ \gamma > 0.99 $ and it is non-monotonic with respect to $ p_{fast} $. (C) $ p_{sr} $ as a function of coefficient of dispersion, $ c_d $. The similarity of the results from protocol IE to that of protocols PF and CD indicates at the equivalence between these three protocols.}
		\label{fig:IE}
	\end{figure}

	In most experiments, it is easier to control the interaction energies of the building blocks (\textit{atoms}) than the rate constant distribution of the generated reaction network. Therefore, our theoretical results will be useful if and only if it can be established that the simplifying assumption of a quasi-random chemical network connectivity is effectively valid for more realistic models in which reaction rate kinetics are determined by underlying physical quantities such as interaction energies between components. We therefore sought next to analyze a ``mechanistic model" in which the activation barriers of the reactions are obtained by assuming a transition state model of the reaction kinetics (Fig.~\ref{fig1}E). The energies of the ground and the transition states are determined by the interaction energies of the atoms (SI), which are allowed to form clusters of up to four members. Therefore, the dispersion of the rate constants can be controlled by changing the interaction energies. Typically, stronger interaction energies correspond to broader distributions of rate constants. Hence, as per our results from protocol CD, we expect to observe exponential growth when the atoms interact strongly with each other. As Fig.~\ref{fig:protocols}D shows, that is indeed the case. Detailed exploration of the interaction energy space shows that this analogy is rigorous (Fig.~\ref{fig:IE}) and these three protocols are potentially equivalent to each other. 
	
	\section{Equivalence of control protocols}
	
	\begin{figure}[htbp]
		\centering
		\includegraphics[width=8.5cm]{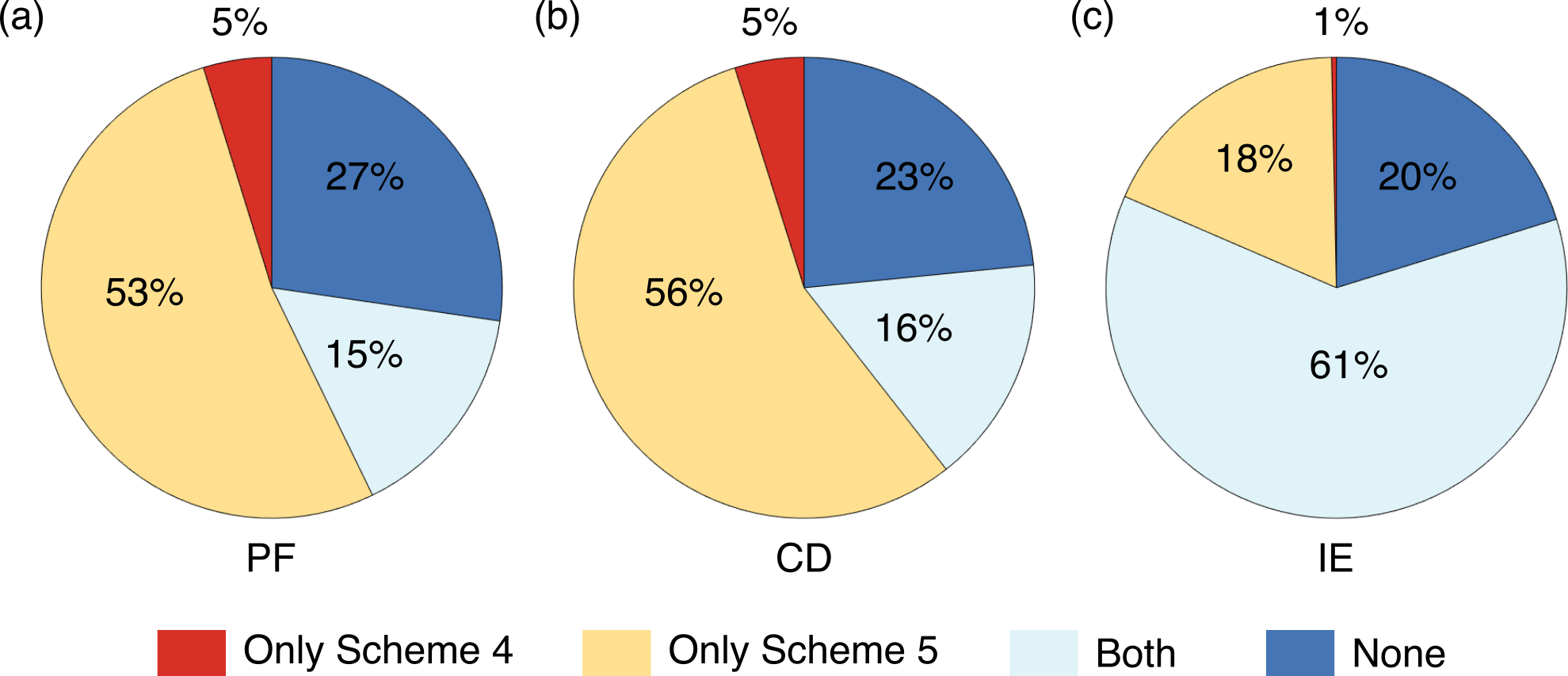}
		\caption{\textbf{Modes of self-replication: } Pie chart showing modes of self-replication in three protocols (a) PF, (b) CD, and (c) IE. Isolated ACCs (schemes 1-3) were absent in all three protocols, and, surprisingly, the dominant mode of self-replication was scheme 5, which contains no isolated ACC. Where no modes of self-replication were registered, it is likely that self-replication happens through other motifs that we have not considered here.}
		\label{fig:schemes}
	\end{figure}
	
	The three protocols described here impose macroscopic control on the reaction kinetics through the rate constants. Although motivated by related physical intuitions, these ensembles of reaction graphs do differ in their microscopic statistics, and it is important to ask whether they ultimately succeed in generating self-replicators for the same underlying topological reasons. Therefore, we sought to understand the modes of self-replication that each of these protocols employs. As Fig.~\ref{fig:schemes} shows, the dominant modes of self-replication are, perhaps surprisingly, scheme 4 and 5 and schemes 1-3 were absent from all three protocols. Although surprising, this result is similar to previous experiments ~\cite{vaidya2012spontaneous}, where isolated ACCs were superseded by cooperative CCs as the main mode of self-replication. Furthermore, the equivalence between the three protocols indicates that the topology of the coupled reaction network plays more important role in determining the transient behavior than the rate constants.
	
\begin{figure}
		\centering
		\includegraphics[width=8.5cm]{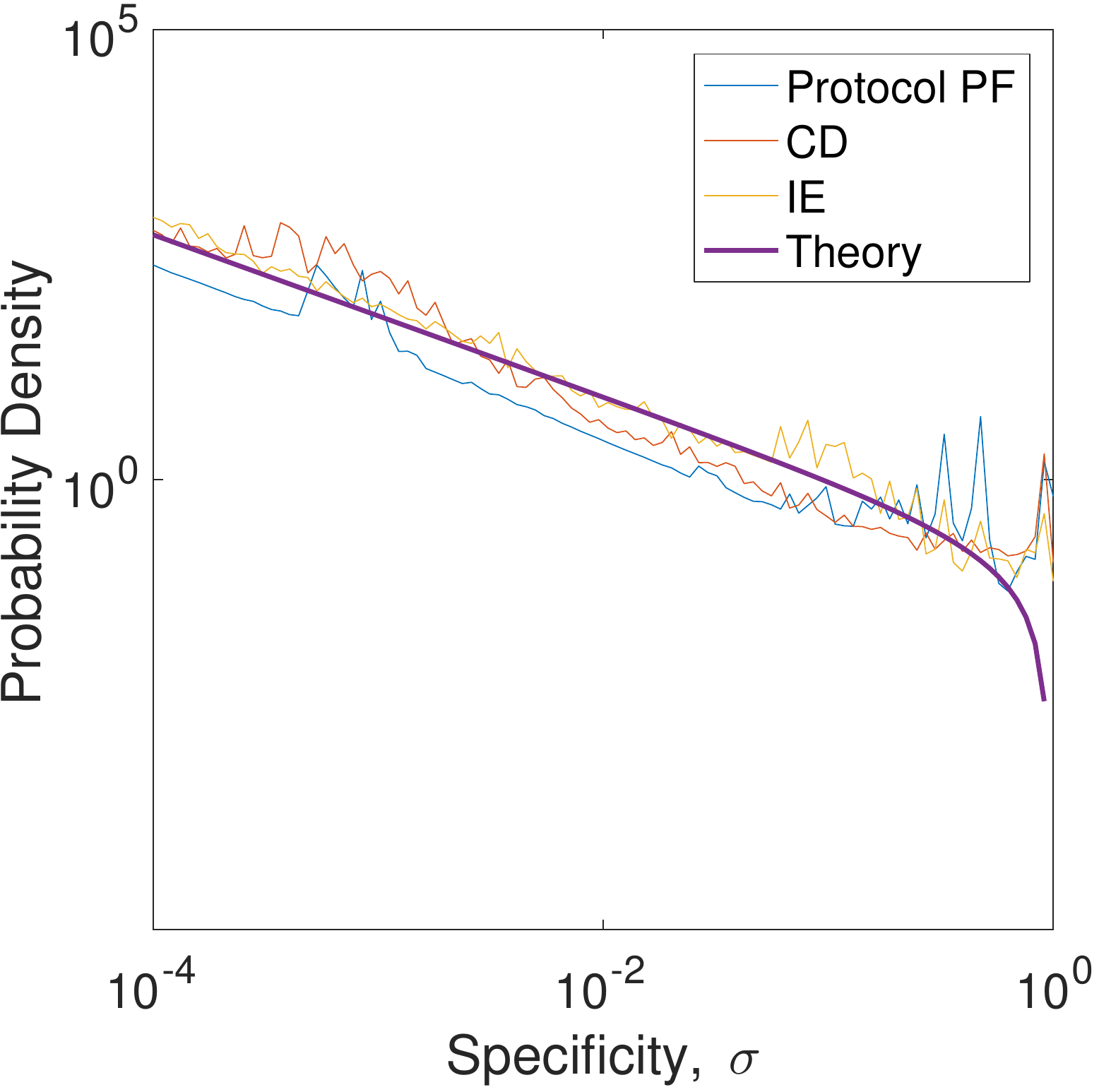}
		\caption{\textbf{PDF of specificity: } PDF of specificity, $\sigma$, from three different protocols and a theoretical estimate assuming that the propensities are distributed as $ \rho_p(x)\sim x^{\nu}exp(-\lambda x) $. The plot shown here corresponds to $ \nu = -0.9 $ and $ \lambda = 0.01 $. Despite the dissimilarity in the choice of the rate constants, the specificity distribution is statistically identical in three different protocols. }
		\label{fig:sigma}
	\end{figure}

	To understand how the choice of the coupled-reaction graph may influence the transient growth behavior, we investigate the outcome of protocol PF under various choices of the underlying coupled-reaction network. The analysis is described in detail in the SI. Here, we describe the set up of the problem. Let's consider a reaction network with $N$ reactions that are coupled with each other with probability $p$. Furthermore, let's assume that a fraction $f_d$ of the $N$ reactions are doubling reactions (reaction of the type: $A + B \rightarrow 2C$). Then, the number of 2-step isolated ACC (scheme 1), scales as:
	\begin{equation}
	n_{1} \sim (N - Nf_d)Nf_dp^2
	\label{eq:n1}
	\end{equation}
	Similarly,
	\begin{eqnarray}
	n_{4} \sim \frac 1 2 (N - Nf_d)^2 Nf_d p^4 \\ 
	n_{5} \sim \frac 1 6 (N - Nf_d)^3 p^4 \label{eq:n45}
	\end{eqnarray}
	It is easy to show from Eq.~\ref{eq:n1}-\ref{eq:n45} that $n_1$ is larger than $n_4$ if $p < \frac{\sqrt{2(1+f_d)}}{N}$, and $n_1$ is larger than $n_5$ if $p < \frac{\sqrt{6f_d(1+2f_d)}}{N}$. Both of these probabilities are incidentally smaller than the average $p$ for our system, which is roughly $\frac{2}{\sqrt N}$. Therefore, purely by numbers, schemes 4 and 5 are more likely over schemes 1-3. However, as we have stated earlier, self-replication occurs only when the specificities of the reactions in a given motif satisfy the required conditions. For schemes 1-3, the specificity of the cycle has to be greater than 0.5 or, on average, the specificities of the reactions comprising the ACCs has to be greater than $\frac{1}{\sqrt{2}} \approx 0.71$. On the other hand, the the conditions for schemes 4 and 5 are much more lenient, as can be verified from Table~\ref{tab:spec}. To estimate the likelihood of meeting these conditions, we estimate the probability distribution of the specificities (SI). Under the assumption that the propensities for various reactions are distributed as $ \rho_p(x) \sim x^{\nu}\exp(-\lambda x) $, the pdf of the specificity $\sigma$, follows the distribution described in Fig.~\ref{fig:sigma}. It is evident from the pdf that one is hardly likely to find reactions with specificities higher than 0.71. On the other hand, one is quite likely to find reactions with specificities less than 0.5, which can satisfy the conditions required for schemes 4 and 5. Furthermore, despite the differences in the choice of the rate constants the specificity distribution from the three protocols are statistically identical to the theoretical approximation. Therefore, structural identity of the coupled reaction graph as well as the statistical similarity of the specificity distribution is the origin of microscopic equivalence between the three different protocols.

	\section{Discussion}
	In this paper, we have developed and investigated a model chemical system, where the constituent chemicals interact with each other through stoichiometric reactions. We have solved this model under three different protocols that impart different levels of macroscopic control over the rate constant distribution of the reactions. We have found out that despite the macroscopic differences, the microscopic kinetics responsible for self-replication is same for all three protocols. In all three protocols, self-replication occurs due to the proliferation of coupled catalytic cycles and not due to isolated autocatalytic cycles, a result similar in spirit to an earlier experiment~\cite{vaidya2012spontaneous}. Furthermore, we have also shown that the criteria for self-replication from the proliferation of an isolated autocatalytic cycle is very different from the criteria for the self-replication of coupled catalytic cycles. In fact, cycle specificity, a well-known metric, can be much less than 0.5 and still the molecules involved can still grow exponentially, in complete violation of the criteria established previously~\cite{szathmary2006origin,king1982recycling}.   
	
	In the light of the results described here, future design of self-replicating systems should focus on developing chemical environment conducive for the proliferation of coupled catalytic cycles as opposed to isolated autocatalytic cycles, since the former can survive even when the reactions are not very specific. Creating such an environment through microscopic tuning of the rate constants, by no means, is easy. However, as we have shown here, it is possible to control coarse features of the chemical network, such as the width of the rate constant distribution, or the interaction energies between the building blocks to achieve the same goal easily.

	Many factors may affect the viability of these design conditions. Firstly, in this paper, we have chosen to report the behavior of the model in a regime in which the supply of the resources is not a limiting factor.   In simulations with limited resources, however, exponential growth can be hindered if the system reaches chemical equilibrium before the onset of the exponential growth, consistent with previous studies~\cite{sievers1994self,szathmary2006origin}. Secondly, we have focused implicitly on the regime of a large and dilute reaction pot where mass-action kinetics applies.  Of course, in any real reactor, the finite total number of particles would lead to small number noisiness in the early emergence and growth of self-replicators that come about from bound states that are initially at low concentration or totally absent.  This means that our results most likely to apply in settings where the components feeding autocatalytic cycles are not themselves difficult to form rapidly from promiscuous reactions among components present in the initial condition.  Finally, it is certain that topological quantities other than $ p_{fast} $ also can play an important role in determining the likelihood of self-replication. For example, the edge degree distribution of the coupled-reaction graph, which is nearly uniform here, is an important determinant of the reaction kinetics.  However, for the purpose of clarity and brevity, we postpone this discussion for the future.
	
	\begin{acknowledgements}
		We would like to thank J. Horowitz, P. Chvykov and other members of the England group for extensive discussion and critical evaluation of the work. Additionally, SS would like to thank B. Chakraborty, P. Mehta, K. Ramola, A. Narayanan, and N. Pal for stimulating discussions that led to the core results of this paper. This work was funded by grants from John Templeton Foundation through grant 55844 and the Gordon and Betty Moore Foundation through grant GBMF4343. JLE is also supported by a Scholar Award (220020476) from the James S. McDonnell Foundation.
	\end{acknowledgements}
	
	\nocite{*}
	\bibliography{pcc.bib}

\begin{thebibliography}{40}%
\makeatletter
\providecommand \@ifxundefined [1]{%
 \@ifx{#1\undefined}
}%
\providecommand \@ifnum [1]{%
 \ifnum #1\expandafter \@firstoftwo
 \else \expandafter \@secondoftwo
 \fi
}%
\providecommand \@ifx [1]{%
 \ifx #1\expandafter \@firstoftwo
 \else \expandafter \@secondoftwo
 \fi
}%
\providecommand \natexlab [1]{#1}%
\providecommand \enquote  [1]{``#1''}%
\providecommand \bibnamefont  [1]{#1}%
\providecommand \bibfnamefont [1]{#1}%
\providecommand \citenamefont [1]{#1}%
\providecommand \href@noop [0]{\@secondoftwo}%
\providecommand \href [0]{\begingroup \@sanitize@url \@href}%
\providecommand \@href[1]{\@@startlink{#1}\@@href}%
\providecommand \@@href[1]{\endgroup#1\@@endlink}%
\providecommand \@sanitize@url [0]{\catcode `\\12\catcode `\$12\catcode
  `\&12\catcode `\#12\catcode `\^12\catcode `\_12\catcode `\%12\relax}%
\providecommand \@@startlink[1]{}%
\providecommand \@@endlink[0]{}%
\providecommand \url  [0]{\begingroup\@sanitize@url \@url }%
\providecommand \@url [1]{\endgroup\@href {#1}{\urlprefix }}%
\providecommand \urlprefix  [0]{URL }%
\providecommand \Eprint [0]{\href }%
\providecommand \doibase [0]{http://dx.doi.org/}%
\providecommand \selectlanguage [0]{\@gobble}%
\providecommand \bibinfo  [0]{\@secondoftwo}%
\providecommand \bibfield  [0]{\@secondoftwo}%
\providecommand \translation [1]{[#1]}%
\providecommand \BibitemOpen [0]{}%
\providecommand \bibitemStop [0]{}%
\providecommand \bibitemNoStop [0]{.\EOS\space}%
\providecommand \EOS [0]{\spacefactor3000\relax}%
\providecommand \BibitemShut  [1]{\csname bibitem#1\endcsname}%
\let\auto@bib@innerbib\@empty
\bibitem [{\citenamefont {Butlerow}(1861)}]{butlerow1861formation}%
  \BibitemOpen
  \bibfield  {author} {\bibinfo {author} {\bibfnamefont {A.}~\bibnamefont
  {Butlerow}},\ }\href@noop {} {\bibfield  {journal} {\bibinfo  {journal} {CR
  Acad. Sci}\ }\textbf {\bibinfo {volume} {53}},\ \bibinfo {pages} {145}
  (\bibinfo {year} {1861})}\BibitemShut {NoStop}%
\bibitem [{\citenamefont {Breslow}(1959)}]{breslow1959mechanism}%
  \BibitemOpen
  \bibfield  {author} {\bibinfo {author} {\bibfnamefont {R.}~\bibnamefont
  {Breslow}},\ }\href@noop {} {\bibfield  {journal} {\bibinfo  {journal}
  {Tetrahedron Letters}\ }\textbf {\bibinfo {volume} {1}},\ \bibinfo {pages}
  {22} (\bibinfo {year} {1959})}\BibitemShut {NoStop}%
\bibitem [{\citenamefont {Dyson}(1982)}]{dyson1982model}%
  \BibitemOpen
  \bibfield  {author} {\bibinfo {author} {\bibfnamefont {F.~J.}\ \bibnamefont
  {Dyson}},\ }\href@noop {} {\bibfield  {journal} {\bibinfo  {journal} {Journal
  of Molecular Evolution}\ }\textbf {\bibinfo {volume} {18}},\ \bibinfo {pages}
  {344} (\bibinfo {year} {1982})}\BibitemShut {NoStop}%
\bibitem [{\citenamefont {Szathm{\'a}ry}(2006)}]{szathmary2006origin}%
  \BibitemOpen
  \bibfield  {author} {\bibinfo {author} {\bibfnamefont {E.}~\bibnamefont
  {Szathm{\'a}ry}},\ }\href@noop {} {\bibfield  {journal} {\bibinfo  {journal}
  {Philosophical Transactions of the Royal Society B: Biological Sciences}\
  }\textbf {\bibinfo {volume} {361}},\ \bibinfo {pages} {1761} (\bibinfo {year}
  {2006})}\BibitemShut {NoStop}%
\bibitem [{\citenamefont {King}(1982)}]{king1982recycling}%
  \BibitemOpen
  \bibfield  {author} {\bibinfo {author} {\bibfnamefont {G.}~\bibnamefont
  {King}},\ }\href@noop {} {\bibfield  {journal} {\bibinfo  {journal}
  {Biosystems}\ }\textbf {\bibinfo {volume} {15}},\ \bibinfo {pages} {89}
  (\bibinfo {year} {1982})}\BibitemShut {NoStop}%
\bibitem [{\citenamefont {Bissette}\ and\ \citenamefont
  {Fletcher}(2013)}]{bissette2013mechanisms}%
  \BibitemOpen
  \bibfield  {author} {\bibinfo {author} {\bibfnamefont {A.~J.}\ \bibnamefont
  {Bissette}}\ and\ \bibinfo {author} {\bibfnamefont {S.~P.}\ \bibnamefont
  {Fletcher}},\ }\href@noop {} {\bibfield  {journal} {\bibinfo  {journal}
  {Angewandte Chemie International Edition}\ }\textbf {\bibinfo {volume}
  {52}},\ \bibinfo {pages} {12800} (\bibinfo {year} {2013})}\BibitemShut
  {NoStop}%
\bibitem [{\citenamefont {Paul}\ and\ \citenamefont
  {Joyce}(2004)}]{paul2004minimal}%
  \BibitemOpen
  \bibfield  {author} {\bibinfo {author} {\bibfnamefont {N.}~\bibnamefont
  {Paul}}\ and\ \bibinfo {author} {\bibfnamefont {G.~F.}\ \bibnamefont
  {Joyce}},\ }\href@noop {} {\bibfield  {journal} {\bibinfo  {journal} {Current
  opinion in chemical biology}\ }\textbf {\bibinfo {volume} {8}},\ \bibinfo
  {pages} {634} (\bibinfo {year} {2004})}\BibitemShut {NoStop}%
\bibitem [{\citenamefont {Carnall}\ \emph {et~al.}(2010)\citenamefont
  {Carnall}, \citenamefont {Waudby}, \citenamefont {Belenguer}, \citenamefont
  {Stuart}, \citenamefont {Peyralans},\ and\ \citenamefont
  {Otto}}]{carnall2010mechanosensitive}%
  \BibitemOpen
  \bibfield  {author} {\bibinfo {author} {\bibfnamefont {J.~M.}\ \bibnamefont
  {Carnall}}, \bibinfo {author} {\bibfnamefont {C.~A.}\ \bibnamefont {Waudby}},
  \bibinfo {author} {\bibfnamefont {A.~M.}\ \bibnamefont {Belenguer}}, \bibinfo
  {author} {\bibfnamefont {M.~C.}\ \bibnamefont {Stuart}}, \bibinfo {author}
  {\bibfnamefont {J.~J.-P.}\ \bibnamefont {Peyralans}}, \ and\ \bibinfo
  {author} {\bibfnamefont {S.}~\bibnamefont {Otto}},\ }\href@noop {} {\bibfield
   {journal} {\bibinfo  {journal} {Science}\ }\textbf {\bibinfo {volume}
  {327}},\ \bibinfo {pages} {1502} (\bibinfo {year} {2010})}\BibitemShut
  {NoStop}%
\bibitem [{\citenamefont {Wang}\ \emph {et~al.}(2011)\citenamefont {Wang},
  \citenamefont {Sha}, \citenamefont {Dreyfus}, \citenamefont {Leunissen},
  \citenamefont {Maass}, \citenamefont {Pine}, \citenamefont {Chaikin},\ and\
  \citenamefont {Seeman}}]{wang2011self}%
  \BibitemOpen
  \bibfield  {author} {\bibinfo {author} {\bibfnamefont {T.}~\bibnamefont
  {Wang}}, \bibinfo {author} {\bibfnamefont {R.}~\bibnamefont {Sha}}, \bibinfo
  {author} {\bibfnamefont {R.}~\bibnamefont {Dreyfus}}, \bibinfo {author}
  {\bibfnamefont {M.~E.}\ \bibnamefont {Leunissen}}, \bibinfo {author}
  {\bibfnamefont {C.}~\bibnamefont {Maass}}, \bibinfo {author} {\bibfnamefont
  {D.~J.}\ \bibnamefont {Pine}}, \bibinfo {author} {\bibfnamefont {P.~M.}\
  \bibnamefont {Chaikin}}, \ and\ \bibinfo {author} {\bibfnamefont {N.~C.}\
  \bibnamefont {Seeman}},\ }\href@noop {} {\bibfield  {journal} {\bibinfo
  {journal} {Nature}\ }\textbf {\bibinfo {volume} {478}},\ \bibinfo {pages}
  {225} (\bibinfo {year} {2011})}\BibitemShut {NoStop}%
\bibitem [{\citenamefont {Vaidya}\ \emph {et~al.}(2012)\citenamefont {Vaidya},
  \citenamefont {Manapat}, \citenamefont {Chen}, \citenamefont {Xulvi-Brunet},
  \citenamefont {Hayden},\ and\ \citenamefont
  {Lehman}}]{vaidya2012spontaneous}%
  \BibitemOpen
  \bibfield  {author} {\bibinfo {author} {\bibfnamefont {N.}~\bibnamefont
  {Vaidya}}, \bibinfo {author} {\bibfnamefont {M.~L.}\ \bibnamefont {Manapat}},
  \bibinfo {author} {\bibfnamefont {I.~A.}\ \bibnamefont {Chen}}, \bibinfo
  {author} {\bibfnamefont {R.}~\bibnamefont {Xulvi-Brunet}}, \bibinfo {author}
  {\bibfnamefont {E.~J.}\ \bibnamefont {Hayden}}, \ and\ \bibinfo {author}
  {\bibfnamefont {N.}~\bibnamefont {Lehman}},\ }\href@noop {} {\bibfield
  {journal} {\bibinfo  {journal} {Nature}\ }\textbf {\bibinfo {volume} {491}},\
  \bibinfo {pages} {72} (\bibinfo {year} {2012})}\BibitemShut {NoStop}%
\bibitem [{\citenamefont {Zeravcic}\ and\ \citenamefont
  {Brenner}(2014)}]{zeravcic2014self}%
  \BibitemOpen
  \bibfield  {author} {\bibinfo {author} {\bibfnamefont {Z.}~\bibnamefont
  {Zeravcic}}\ and\ \bibinfo {author} {\bibfnamefont {M.~P.}\ \bibnamefont
  {Brenner}},\ }\href@noop {} {\bibfield  {journal} {\bibinfo  {journal}
  {Proceedings of the National Academy of Sciences}\ }\textbf {\bibinfo
  {volume} {111}},\ \bibinfo {pages} {1748} (\bibinfo {year}
  {2014})}\BibitemShut {NoStop}%
\bibitem [{\citenamefont {Sadownik}\ \emph {et~al.}(2016)\citenamefont
  {Sadownik}, \citenamefont {Mattia}, \citenamefont {Nowak},\ and\
  \citenamefont {Otto}}]{sadownik2016diversification}%
  \BibitemOpen
  \bibfield  {author} {\bibinfo {author} {\bibfnamefont {J.~W.}\ \bibnamefont
  {Sadownik}}, \bibinfo {author} {\bibfnamefont {E.}~\bibnamefont {Mattia}},
  \bibinfo {author} {\bibfnamefont {P.}~\bibnamefont {Nowak}}, \ and\ \bibinfo
  {author} {\bibfnamefont {S.}~\bibnamefont {Otto}},\ }\href@noop {} {\bibfield
   {journal} {\bibinfo  {journal} {Nature chemistry}\ } (\bibinfo {year}
  {2016})}\BibitemShut {NoStop}%
\bibitem [{\citenamefont {{\v{S}}ari{\'c}}\ \emph {et~al.}(2016)\citenamefont
  {{\v{S}}ari{\'c}}, \citenamefont {Buell}, \citenamefont {Meisl},
  \citenamefont {Michaels}, \citenamefont {Dobson}, \citenamefont {Linse},
  \citenamefont {Knowles},\ and\ \citenamefont {Frenkel}}]{vsaric2016physical}%
  \BibitemOpen
  \bibfield  {author} {\bibinfo {author} {\bibfnamefont {A.}~\bibnamefont
  {{\v{S}}ari{\'c}}}, \bibinfo {author} {\bibfnamefont {A.~K.}\ \bibnamefont
  {Buell}}, \bibinfo {author} {\bibfnamefont {G.}~\bibnamefont {Meisl}},
  \bibinfo {author} {\bibfnamefont {T.~C.}\ \bibnamefont {Michaels}}, \bibinfo
  {author} {\bibfnamefont {C.~M.}\ \bibnamefont {Dobson}}, \bibinfo {author}
  {\bibfnamefont {S.}~\bibnamefont {Linse}}, \bibinfo {author} {\bibfnamefont
  {T.~P.}\ \bibnamefont {Knowles}}, \ and\ \bibinfo {author} {\bibfnamefont
  {D.}~\bibnamefont {Frenkel}},\ }\href@noop {} {\bibfield  {journal} {\bibinfo
   {journal} {Nature Physics}\ } (\bibinfo {year} {2016})}\BibitemShut
  {NoStop}%
\bibitem [{\citenamefont {Barenholz}\ \emph {et~al.}(2017)\citenamefont
  {Barenholz}, \citenamefont {Davidi}, \citenamefont {Reznik}, \citenamefont
  {Bar-On}, \citenamefont {Antonovsky}, \citenamefont {Noor},\ and\
  \citenamefont {Milo}}]{barenholz2017design}%
  \BibitemOpen
  \bibfield  {author} {\bibinfo {author} {\bibfnamefont {U.}~\bibnamefont
  {Barenholz}}, \bibinfo {author} {\bibfnamefont {D.}~\bibnamefont {Davidi}},
  \bibinfo {author} {\bibfnamefont {E.}~\bibnamefont {Reznik}}, \bibinfo
  {author} {\bibfnamefont {Y.}~\bibnamefont {Bar-On}}, \bibinfo {author}
  {\bibfnamefont {N.}~\bibnamefont {Antonovsky}}, \bibinfo {author}
  {\bibfnamefont {E.}~\bibnamefont {Noor}}, \ and\ \bibinfo {author}
  {\bibfnamefont {R.}~\bibnamefont {Milo}},\ }\href@noop {} {\bibfield
  {journal} {\bibinfo  {journal} {eLife}\ }\textbf {\bibinfo {volume} {6}},\
  \bibinfo {pages} {e20667} (\bibinfo {year} {2017})}\BibitemShut {NoStop}%
\bibitem [{\citenamefont {Zwicker}\ \emph {et~al.}(2017)\citenamefont
  {Zwicker}, \citenamefont {Seyboldt}, \citenamefont {Weber}, \citenamefont
  {Hyman},\ and\ \citenamefont {J{\"u}licher}}]{zwicker2017growth}%
  \BibitemOpen
  \bibfield  {author} {\bibinfo {author} {\bibfnamefont {D.}~\bibnamefont
  {Zwicker}}, \bibinfo {author} {\bibfnamefont {R.}~\bibnamefont {Seyboldt}},
  \bibinfo {author} {\bibfnamefont {C.~A.}\ \bibnamefont {Weber}}, \bibinfo
  {author} {\bibfnamefont {A.~A.}\ \bibnamefont {Hyman}}, \ and\ \bibinfo
  {author} {\bibfnamefont {F.}~\bibnamefont {J{\"u}licher}},\ }\href@noop {}
  {\bibfield  {journal} {\bibinfo  {journal} {Nature Physics}\ }\textbf
  {\bibinfo {volume} {13}},\ \bibinfo {pages} {408} (\bibinfo {year}
  {2017})}\BibitemShut {NoStop}%
\bibitem [{\citenamefont {England}(2013)}]{england2013statistical}%
  \BibitemOpen
  \bibfield  {author} {\bibinfo {author} {\bibfnamefont {J.~L.}\ \bibnamefont
  {England}},\ }\href@noop {} {\bibfield  {journal} {\bibinfo  {journal} {The
  Journal of chemical physics}\ }\textbf {\bibinfo {volume} {139}},\ \bibinfo
  {pages} {09B623\_1} (\bibinfo {year} {2013})}\BibitemShut {NoStop}%
\bibitem [{\citenamefont {Perunov}\ \emph {et~al.}(2016)\citenamefont
  {Perunov}, \citenamefont {Marsland},\ and\ \citenamefont
  {England}}]{perunov2016statistical}%
  \BibitemOpen
  \bibfield  {author} {\bibinfo {author} {\bibfnamefont {N.}~\bibnamefont
  {Perunov}}, \bibinfo {author} {\bibfnamefont {R.~A.}\ \bibnamefont
  {Marsland}}, \ and\ \bibinfo {author} {\bibfnamefont {J.~L.}\ \bibnamefont
  {England}},\ }\href@noop {} {\bibfield  {journal} {\bibinfo  {journal}
  {Physical Review X}\ }\textbf {\bibinfo {volume} {6}},\ \bibinfo {pages}
  {021036} (\bibinfo {year} {2016})}\BibitemShut {NoStop}%
\bibitem [{\citenamefont {Kauffman}(1986)}]{kauffman1986autocatalytic}%
  \BibitemOpen
  \bibfield  {author} {\bibinfo {author} {\bibfnamefont {S.~A.}\ \bibnamefont
  {Kauffman}},\ }\href@noop {} {\bibfield  {journal} {\bibinfo  {journal}
  {Journal of theoretical biology}\ }\textbf {\bibinfo {volume} {119}},\
  \bibinfo {pages} {1} (\bibinfo {year} {1986})}\BibitemShut {NoStop}%
\bibitem [{\citenamefont {Jain}\ and\ \citenamefont
  {Krishna}(2001)}]{jain2001model}%
  \BibitemOpen
  \bibfield  {author} {\bibinfo {author} {\bibfnamefont {S.}~\bibnamefont
  {Jain}}\ and\ \bibinfo {author} {\bibfnamefont {S.}~\bibnamefont {Krishna}},\
  }\href@noop {} {\bibfield  {journal} {\bibinfo  {journal} {Proceedings of the
  National Academy of Sciences}\ }\textbf {\bibinfo {volume} {98}},\ \bibinfo
  {pages} {543} (\bibinfo {year} {2001})}\BibitemShut {NoStop}%
\bibitem [{\citenamefont {Zeravcic}\ and\ \citenamefont
  {Brenner}(2017{\natexlab{a}})}]{zeravcic2017spontaneous}%
  \BibitemOpen
  \bibfield  {author} {\bibinfo {author} {\bibfnamefont {Z.}~\bibnamefont
  {Zeravcic}}\ and\ \bibinfo {author} {\bibfnamefont {M.~P.}\ \bibnamefont
  {Brenner}},\ }\href@noop {} {\bibfield  {journal} {\bibinfo  {journal}
  {Proceedings of the National Academy of Sciences}\ ,\ \bibinfo {pages}
  {201611959}} (\bibinfo {year} {2017}{\natexlab{a}})}\BibitemShut {NoStop}%
\bibitem [{\citenamefont {Wang}\ \emph {et~al.}(2012)\citenamefont {Wang},
  \citenamefont {Wang}, \citenamefont {Breed}, \citenamefont {Manoharan},
  \citenamefont {Feng}, \citenamefont {Hollingsworth}, \citenamefont {Weck},\
  and\ \citenamefont {Pine}}]{wang2012colloids}%
  \BibitemOpen
  \bibfield  {author} {\bibinfo {author} {\bibfnamefont {Y.}~\bibnamefont
  {Wang}}, \bibinfo {author} {\bibfnamefont {Y.}~\bibnamefont {Wang}}, \bibinfo
  {author} {\bibfnamefont {D.~R.}\ \bibnamefont {Breed}}, \bibinfo {author}
  {\bibfnamefont {V.~N.}\ \bibnamefont {Manoharan}}, \bibinfo {author}
  {\bibfnamefont {L.}~\bibnamefont {Feng}}, \bibinfo {author} {\bibfnamefont
  {A.~D.}\ \bibnamefont {Hollingsworth}}, \bibinfo {author} {\bibfnamefont
  {M.}~\bibnamefont {Weck}}, \ and\ \bibinfo {author} {\bibfnamefont {D.~J.}\
  \bibnamefont {Pine}},\ }\href@noop {} {\  (\bibinfo {year}
  {2012})}\BibitemShut {NoStop}%
\bibitem [{\citenamefont {Zhang}\ \emph {et~al.}(2014)\citenamefont {Zhang},
  \citenamefont {Dempster},\ and\ \citenamefont {de~la Cruz}}]{zhang2014self}%
  \BibitemOpen
  \bibfield  {author} {\bibinfo {author} {\bibfnamefont {R.}~\bibnamefont
  {Zhang}}, \bibinfo {author} {\bibfnamefont {J.~M.}\ \bibnamefont {Dempster}},
  \ and\ \bibinfo {author} {\bibfnamefont {M.~O.}\ \bibnamefont {de~la Cruz}},\
  }\href@noop {} {\bibfield  {journal} {\bibinfo  {journal} {Soft matter}\
  }\textbf {\bibinfo {volume} {10}},\ \bibinfo {pages} {1315} (\bibinfo {year}
  {2014})}\BibitemShut {NoStop}%
\bibitem [{\citenamefont {Sievers}\ and\ \citenamefont
  {Von~Kiedrowski}(1994)}]{sievers1994self}%
  \BibitemOpen
  \bibfield  {author} {\bibinfo {author} {\bibfnamefont {D.}~\bibnamefont
  {Sievers}}\ and\ \bibinfo {author} {\bibfnamefont {G.}~\bibnamefont
  {Von~Kiedrowski}},\ }\href@noop {} {\bibfield  {journal} {\bibinfo  {journal}
  {Nature}\ }\textbf {\bibinfo {volume} {369}},\ \bibinfo {pages} {221}
  (\bibinfo {year} {1994})}\BibitemShut {NoStop}%
\bibitem [{\citenamefont {Davidi}\ \emph {et~al.}(2016)\citenamefont {Davidi},
  \citenamefont {Noor}, \citenamefont {Liebermeister}, \citenamefont
  {Bar-Even}, \citenamefont {Flamholz}, \citenamefont {Tummler}, \citenamefont
  {Barenholz}, \citenamefont {Goldenfeld}, \citenamefont {Shlomi},\ and\
  \citenamefont {Milo}}]{davidi2016global}%
  \BibitemOpen
  \bibfield  {author} {\bibinfo {author} {\bibfnamefont {D.}~\bibnamefont
  {Davidi}}, \bibinfo {author} {\bibfnamefont {E.}~\bibnamefont {Noor}},
  \bibinfo {author} {\bibfnamefont {W.}~\bibnamefont {Liebermeister}}, \bibinfo
  {author} {\bibfnamefont {A.}~\bibnamefont {Bar-Even}}, \bibinfo {author}
  {\bibfnamefont {A.}~\bibnamefont {Flamholz}}, \bibinfo {author}
  {\bibfnamefont {K.}~\bibnamefont {Tummler}}, \bibinfo {author} {\bibfnamefont
  {U.}~\bibnamefont {Barenholz}}, \bibinfo {author} {\bibfnamefont
  {M.}~\bibnamefont {Goldenfeld}}, \bibinfo {author} {\bibfnamefont
  {T.}~\bibnamefont {Shlomi}}, \ and\ \bibinfo {author} {\bibfnamefont
  {R.}~\bibnamefont {Milo}},\ }\href@noop {} {\bibfield  {journal} {\bibinfo
  {journal} {Proceedings of the National Academy of Sciences}\ }\textbf
  {\bibinfo {volume} {113}},\ \bibinfo {pages} {3401} (\bibinfo {year}
  {2016})}\BibitemShut {NoStop}%
\bibitem [{\citenamefont {Eigen}\ and\ \citenamefont
  {Schuster}(2012)}]{eigen2012hypercycle}%
  \BibitemOpen
  \bibfield  {author} {\bibinfo {author} {\bibfnamefont {M.}~\bibnamefont
  {Eigen}}\ and\ \bibinfo {author} {\bibfnamefont {P.}~\bibnamefont
  {Schuster}},\ }\href@noop {} {\emph {\bibinfo {title} {The hypercycle: a
  principle of natural self-organization}}}\ (\bibinfo  {publisher} {Springer
  Science \& Business Media},\ \bibinfo {year} {2012})\BibitemShut {NoStop}%
\bibitem [{\citenamefont {Epstein}\ and\ \citenamefont
  {Pojman}(1998)}]{epstein1998introduction}%
  \BibitemOpen
  \bibfield  {author} {\bibinfo {author} {\bibfnamefont {I.~R.}\ \bibnamefont
  {Epstein}}\ and\ \bibinfo {author} {\bibfnamefont {J.~A.}\ \bibnamefont
  {Pojman}},\ }\href@noop {} {\enquote {\bibinfo {title} {An introduction to
  nonlinear chemical dynamics: oscillations, waves, patterns, and chaos},}\ }
  (\bibinfo {year} {1998})\BibitemShut {NoStop}%
\bibitem [{\citenamefont {Griffith}(1967)}]{griffith1967self}%
  \BibitemOpen
  \bibfield  {author} {\bibinfo {author} {\bibfnamefont {J.~S.}\ \bibnamefont
  {Griffith}},\ }\href@noop {} {\bibfield  {journal} {\bibinfo  {journal}
  {Nature}\ }\textbf {\bibinfo {volume} {215}},\ \bibinfo {pages} {1043}
  (\bibinfo {year} {1967})}\BibitemShut {NoStop}%
\bibitem [{\citenamefont {Imparato}\ \emph {et~al.}(2007)\citenamefont
  {Imparato}, \citenamefont {Peliti}, \citenamefont {Pesce}, \citenamefont
  {Rusciano},\ and\ \citenamefont {Sasso}}]{imparato2007work}%
  \BibitemOpen
  \bibfield  {author} {\bibinfo {author} {\bibfnamefont {A.}~\bibnamefont
  {Imparato}}, \bibinfo {author} {\bibfnamefont {L.}~\bibnamefont {Peliti}},
  \bibinfo {author} {\bibfnamefont {G.}~\bibnamefont {Pesce}}, \bibinfo
  {author} {\bibfnamefont {G.}~\bibnamefont {Rusciano}}, \ and\ \bibinfo
  {author} {\bibfnamefont {A.}~\bibnamefont {Sasso}},\ }\href@noop {}
  {\bibfield  {journal} {\bibinfo  {journal} {Physical Review E}\ }\textbf
  {\bibinfo {volume} {76}},\ \bibinfo {pages} {050101} (\bibinfo {year}
  {2007})}\BibitemShut {NoStop}%
\bibitem [{\citenamefont {Jucker}\ and\ \citenamefont
  {Walker}(2013)}]{jucker2013self}%
  \BibitemOpen
  \bibfield  {author} {\bibinfo {author} {\bibfnamefont {M.}~\bibnamefont
  {Jucker}}\ and\ \bibinfo {author} {\bibfnamefont {L.~C.}\ \bibnamefont
  {Walker}},\ }\href@noop {} {\bibfield  {journal} {\bibinfo  {journal}
  {Nature}\ }\textbf {\bibinfo {volume} {501}},\ \bibinfo {pages} {45}
  (\bibinfo {year} {2013})}\BibitemShut {NoStop}%
\bibitem [{\citenamefont {Kruger}\ \emph {et~al.}(1982)\citenamefont {Kruger},
  \citenamefont {Grabowski}, \citenamefont {Zaug}, \citenamefont {Sands},
  \citenamefont {Gottschling},\ and\ \citenamefont {Cech}}]{kruger1982self}%
  \BibitemOpen
  \bibfield  {author} {\bibinfo {author} {\bibfnamefont {K.}~\bibnamefont
  {Kruger}}, \bibinfo {author} {\bibfnamefont {P.~J.}\ \bibnamefont
  {Grabowski}}, \bibinfo {author} {\bibfnamefont {A.~J.}\ \bibnamefont {Zaug}},
  \bibinfo {author} {\bibfnamefont {J.}~\bibnamefont {Sands}}, \bibinfo
  {author} {\bibfnamefont {D.~E.}\ \bibnamefont {Gottschling}}, \ and\ \bibinfo
  {author} {\bibfnamefont {T.~R.}\ \bibnamefont {Cech}},\ }\href@noop {}
  {\bibfield  {journal} {\bibinfo  {journal} {cell}\ }\textbf {\bibinfo
  {volume} {31}},\ \bibinfo {pages} {147} (\bibinfo {year} {1982})}\BibitemShut
  {NoStop}%
\bibitem [{\citenamefont {Lee}\ \emph {et~al.}(1996)\citenamefont {Lee},
  \citenamefont {Granja}, \citenamefont {Martinez}, \citenamefont {Severin},\
  and\ \citenamefont {Ghadiri}}]{lee1996self}%
  \BibitemOpen
  \bibfield  {author} {\bibinfo {author} {\bibfnamefont {D.~H.}\ \bibnamefont
  {Lee}}, \bibinfo {author} {\bibfnamefont {J.~R.}\ \bibnamefont {Granja}},
  \bibinfo {author} {\bibfnamefont {J.~A.}\ \bibnamefont {Martinez}}, \bibinfo
  {author} {\bibfnamefont {K.}~\bibnamefont {Severin}}, \ and\ \bibinfo
  {author} {\bibfnamefont {M.~R.}\ \bibnamefont {Ghadiri}},\ }\href@noop {}
  {\bibfield  {journal} {\bibinfo  {journal} {Nature}\ }\textbf {\bibinfo
  {volume} {382}},\ \bibinfo {pages} {525} (\bibinfo {year}
  {1996})}\BibitemShut {NoStop}%
\bibitem [{\citenamefont {Lincoln}\ and\ \citenamefont
  {Joyce}(2009)}]{lincoln2009self}%
  \BibitemOpen
  \bibfield  {author} {\bibinfo {author} {\bibfnamefont {T.~A.}\ \bibnamefont
  {Lincoln}}\ and\ \bibinfo {author} {\bibfnamefont {G.~F.}\ \bibnamefont
  {Joyce}},\ }\href@noop {} {\bibfield  {journal} {\bibinfo  {journal}
  {Science}\ }\textbf {\bibinfo {volume} {323}},\ \bibinfo {pages} {1229}
  (\bibinfo {year} {2009})}\BibitemShut {NoStop}%
\bibitem [{\citenamefont {Miller}\ and\ \citenamefont
  {Urey}(1959)}]{miller1959organic}%
  \BibitemOpen
  \bibfield  {author} {\bibinfo {author} {\bibfnamefont {S.~L.}\ \bibnamefont
  {Miller}}\ and\ \bibinfo {author} {\bibfnamefont {H.~C.}\ \bibnamefont
  {Urey}},\ }\href@noop {} {\bibfield  {journal} {\bibinfo  {journal}
  {Science}\ }\textbf {\bibinfo {volume} {130}},\ \bibinfo {pages} {245}
  (\bibinfo {year} {1959})}\BibitemShut {NoStop}%
\bibitem [{\citenamefont {Orgel}(1995)}]{orgel1995unnatural}%
  \BibitemOpen
  \bibfield  {author} {\bibinfo {author} {\bibfnamefont {L.~E.}\ \bibnamefont
  {Orgel}},\ }\href@noop {} {\bibfield  {journal} {\bibinfo  {journal}
  {Accounts of chemical research}\ }\textbf {\bibinfo {volume} {28}},\ \bibinfo
  {pages} {109} (\bibinfo {year} {1995})}\BibitemShut {NoStop}%
\bibitem [{\citenamefont {Polettini}\ and\ \citenamefont
  {Esposito}(2014)}]{polettini2014irreversible}%
  \BibitemOpen
  \bibfield  {author} {\bibinfo {author} {\bibfnamefont {M.}~\bibnamefont
  {Polettini}}\ and\ \bibinfo {author} {\bibfnamefont {M.}~\bibnamefont
  {Esposito}},\ }\href@noop {} {\bibfield  {journal} {\bibinfo  {journal} {The
  Journal of chemical physics}\ }\textbf {\bibinfo {volume} {141}},\ \bibinfo
  {pages} {07B610\_1} (\bibinfo {year} {2014})}\BibitemShut {NoStop}%
\bibitem [{\citenamefont {Richter}\ and\ \citenamefont
  {Engelbrecht}(2014)}]{richter2014recent}%
  \BibitemOpen
  \bibfield  {author} {\bibinfo {author} {\bibfnamefont {H.}~\bibnamefont
  {Richter}}\ and\ \bibinfo {author} {\bibfnamefont {A.}~\bibnamefont
  {Engelbrecht}},\ }\href@noop {} {\emph {\bibinfo {title} {Recent advances in
  the theory and application of fitness landscapes}}}\ (\bibinfo  {publisher}
  {Springer},\ \bibinfo {year} {2014})\BibitemShut {NoStop}%
\bibitem [{\citenamefont {Semenov}\ \emph {et~al.}(2016)\citenamefont
  {Semenov}, \citenamefont {Kraft}, \citenamefont {Ainla}, \citenamefont
  {Zhao}, \citenamefont {Baghbanzadeh}, \citenamefont {Campbell}, \citenamefont
  {Kang}, \citenamefont {Fox},\ and\ \citenamefont
  {Whitesides}}]{semenov2016autocatalytic}%
  \BibitemOpen
  \bibfield  {author} {\bibinfo {author} {\bibfnamefont {S.~N.}\ \bibnamefont
  {Semenov}}, \bibinfo {author} {\bibfnamefont {L.~J.}\ \bibnamefont {Kraft}},
  \bibinfo {author} {\bibfnamefont {A.}~\bibnamefont {Ainla}}, \bibinfo
  {author} {\bibfnamefont {M.}~\bibnamefont {Zhao}}, \bibinfo {author}
  {\bibfnamefont {M.}~\bibnamefont {Baghbanzadeh}}, \bibinfo {author}
  {\bibfnamefont {V.~E.}\ \bibnamefont {Campbell}}, \bibinfo {author}
  {\bibfnamefont {K.}~\bibnamefont {Kang}}, \bibinfo {author} {\bibfnamefont
  {J.~M.}\ \bibnamefont {Fox}}, \ and\ \bibinfo {author} {\bibfnamefont
  {G.~M.}\ \bibnamefont {Whitesides}},\ }\href@noop {} {\bibfield  {journal}
  {\bibinfo  {journal} {Nature}\ }\textbf {\bibinfo {volume} {537}},\ \bibinfo
  {pages} {656} (\bibinfo {year} {2016})}\BibitemShut {NoStop}%
\bibitem [{\citenamefont {Szostak}\ \emph {et~al.}(2001)\citenamefont
  {Szostak}, \citenamefont {Bartel},\ and\ \citenamefont
  {Luisi}}]{szostak2001synthesizing}%
  \BibitemOpen
  \bibfield  {author} {\bibinfo {author} {\bibfnamefont {J.~W.}\ \bibnamefont
  {Szostak}}, \bibinfo {author} {\bibfnamefont {D.~P.}\ \bibnamefont {Bartel}},
  \ and\ \bibinfo {author} {\bibfnamefont {P.~L.}\ \bibnamefont {Luisi}},\
  }\href@noop {} {\bibfield  {journal} {\bibinfo  {journal} {Nature}\ }\textbf
  {\bibinfo {volume} {409}},\ \bibinfo {pages} {387} (\bibinfo {year}
  {2001})}\BibitemShut {NoStop}%
\bibitem [{\citenamefont {Zhabotinsky}(1964)}]{zhabotinsky1964periodical}%
  \BibitemOpen
  \bibfield  {author} {\bibinfo {author} {\bibfnamefont {A.~M.}\ \bibnamefont
  {Zhabotinsky}},\ }\href@noop {} {\bibfield  {journal} {\bibinfo  {journal}
  {Biofizika}\ }\textbf {\bibinfo {volume} {9}},\ \bibinfo {pages} {306}
  (\bibinfo {year} {1964})}\BibitemShut {NoStop}%
\bibitem [{\citenamefont {Zeravcic}\ and\ \citenamefont
  {Brenner}(2017{\natexlab{b}})}]{Zeravcic2017}%
  \BibitemOpen
  \bibfield  {author} {\bibinfo {author} {\bibfnamefont {Z.}~\bibnamefont
  {Zeravcic}}\ and\ \bibinfo {author} {\bibfnamefont {M.~P.}\ \bibnamefont
  {Brenner}},\ }\href@noop {} {\bibfield  {journal} {\bibinfo  {journal}
  {Proceedings of the National Academy of Sciences}\ ,\ \bibinfo {pages}
  {201611959}} (\bibinfo {year} {2017}{\natexlab{b}})}\BibitemShut {NoStop}%
\end{thebibliography}%
	
	\appendix
	\section{Materials and Methods}
	\subsection{Numerical solution of differential equations}
	We solved the systems of reactions assuming mass action kinetics. The concentrations of $ B $ and $ G $ were kept constant at 1, whereas the other molecules were initialized with concentration 0. We solved the resultant systems of differential equation with ODE23tb, a stiff solver in matlab. The simulations were run until the system reached chemical equilibrium. Due to the stiffness of the differential equations, the solution sometimes failed to reach chemical equilibria during the runtime of the code, but it did not affect the growth regime. Hence, all the results reported here are unaffected by this limitation of the numerical algorithm.
	
	\subsection{Useful thermodynamic quantities:}
	\paragraph{Propensity or rate} is the product of the rate constant of a reaction and the concentration of the reactants raised to appropriate power. For example, for a reaction: $ A + B -> C + D $ with rate constant $ k_+ $, and obeying mass action kinetics, the propensity is $ k_+[A][B] $, where $ [X] $ denotes the concentration of the reactant $ X $. 
	
	\paragraph{Chemical Current: } Denoted $ J $, it is the difference between the propensities of the forward and reverse reactions of a reversible reaction. For example, for the reaction described earlier, $ J = k_+[A][B] - k_-[C][D] $.   
	
	\subsection{Specificity} Denoted here as $ \sigma $. The specificity is the ratio of the propensity of a given reaction to sum of the propensities of all reactions that consume the resources required for the given reaction, including itself~\cite{szathmary2006origin,king1982recycling}. Mathematically, if $ \pi_i $ is the propensity of reaction $ i $, then 
	\begin{equation}
	\sigma = \frac {\pi_i}{\pi_i + \sum_{j\in \mathcal C} \pi_j}
	\end{equation}
	, where $ \mathcal C $ is the set of all parasitic reactions that consume the resources required for reaction $ i $. $ C = |\mathcal C| $ is the number of such parasitic reactions. The cycle specificity is the product of the specificities of the reactions in the cycle. 
	
	In previous works~\cite{szathmary2006origin,king1982recycling}, specificity was defined strictly for completely irreversible reactions. Therefore, its definition has to be modified for our system, where the reactions are reversible. We have found out that if the chemical current for a reaction is negative it does not contribute to the calculation of the specificity. Therefore, to measure specificity, we have only used reactions whose chemical current is positive. Furthermore, often the concentrations of molecules span several orders of magnitude. Some of them may reach very close to their equilibrium concentration much before other molecules. Under such condition, the concentration of these molecules are unaffected by the consumption of various reactions. As a result, we have ignored any parasitic reaction that consume these molecules from our calculation of specificity.  
	
	\subsection{Growth exponent}
	
	At any given instant, $ t $, the instantaneous growth rate of the concentration, $ dc(t)/dt $, is a simple algebraic function of the concentration, $ c(t) $. Formally,
	\begin{equation}
	\frac {dc}{dt} = rc^\gamma,
	\end{equation}
	where $ \gamma $ is the \textbf{growth exponent} and $ r $ is a proportionality constant. For exponential growth $ \gamma = 1$, for power law (subexponential) growth $ 0 < \gamma < 1 $, and for linear growth $ \gamma = 0 $.  When the concentration grows exponentially ($ \gamma = 1 $), $ r $ is equal to the exponential growth rate constant.
	
	In a typical timeseries, $ \gamma $ varies with time. Therefore, to assess the occurence of exponential growth, in this paper, we measure and report only the maximum value of $ \gamma $ over a timeseries, also referred to as $ \gamma $.
	
	\subsection{Estimate of $ p_{fast} $} 
	
	To estimate $ p_{fast} $ from the time series of the molecular concentrations, we find the fraction of reactions whose propensities are within $ 10\% $ of the propensity of the reaction with fastest propensity. This is a heuristic definition and we have found out that the result does not change as long as it varies between $1-20\%$. For smaller values, the quantitative result changes, but qualitative result remains the same.  
	
	\subsection{Random sampling}
	We sampled 100 different configurations for each random activation barrier ensemble.
	To estimate $ p_{sr} $ in Fig,~\ref{fig:IE}, we binned the scatter plot into different parameter values ($ c_d $ or $ p_{fast} $). Any bins with less than five datapoints were ignored.

\end{document}